\documentclass[12pt]{iopart}
\usepackage{amsfonts}
\usepackage{amssymb}
\usepackage{graphicx}
\usepackage{epsfig}
\usepackage{iopams}
\usepackage{hyperref}
\usepackage{cite}
\usepackage{color}

\begin{document}

\title{Stable dynamic helix state in the nonintegrable XXZ Heisenberg model}
\author{G. Zhang and Z. Song{$^{\dagger}$}}

\ead{songtc@nankai.edu.cn}

\address{School of Physics, Nankai University, Tianjin 300071, China}
\begin{abstract}
{We investigate the influence of external fields on the stability of spin
helix states in an XXZ Heisenberg model. Exact diagonalization on a finite
system shows that random transverse fields in the $x$ and $y$ directions
drive the transition from integrability to nonintegrability. In such a
system, the helix state can be regarded as a quantum scar. Simultaneously,
the presence of uniform $z$ field enables the helix state to better maintain
its dynamical nature, allowing for a clearer understanding of its
evolutionary behavior. However, the entanglement entropy reveals that
irrespective of the presence of a uniform $z$ field, as long as the system
remains chaotic, the scar extent of the helix state shows no significant
variation.}
\end{abstract}

\pacs{05.45.Mt, 05.70.-a, 67.57.Lm, 03.67.-a}

\noindent\textit{Keywords}: Quantum Scar, Quantum Thermalization, Spin
dynamics, Quantum information
\maketitle

\section{Introduction}

The thermalization of quantum systems is one of the main obstacles for
quantum simulation and quantum information processing, because the process
of thermalization always eventually destroys the information of an initial
state. It is expected that there are exceptional cases where some special
slowly thermalizing initial states retain the memory over longer periods of
time. Specifically, it is well established that some nonintegrable systems
can fail to thermalize due to rare nonthermal eigenstates called quantum
many-body scars (QMBS) \cite%
{Shiraishi2017,Moudgalya2018,Moudgalya20182,Khemani2019,Ho2019,Shibata2020,McClarty2020,Richter2022,Jeyaretnam2021,Turner2018,Turner20182,Shiraishi2019,Lin2019,Choi2019,Khemani2020,Dooley2020,Dooley2021,Schecter2019,Iversen2023}%
. These nonthermal states are typically excited ones and span a subspace, in
which any initial states do not thermalize and can return periodically. The
main task in this field is finding scars in a variety of nonintegrable
many-body systems. On the other hand, a modern challenge in condensed matter
physics\ is searching for long-lived non-thermal excited states that
possess\ macroscopic long-range order. In contrast to the ground state,
which is based on a cooling down mechanism, the preparation of\ such states
can be obtained via\ dynamic process.

Atomic systems are an excellent test-beds for quantum simulator in
experiments \cite%
{zhang2017observation,bernien2017probing,barends2015digital,davis2020protecting,signoles2021glassy,trotzky2008time,gross2017quantum}%
, stimulating theoretical studies on the dynamics of quantum spin systems.
These studies not only capture the properties of many artificial systems,
but also provide tractable theoretical examples for understanding
fundamental concepts in physics. As a paradigmatic quantum spin model, the
Heisenberg XXZ model exhibits strong correlations and its dynamical
properties attract attention from both the condensed matter physics and
mathematical-physics communities \cite%
{Keselman2020,Bera2020,Chauhan2020,Babenko2021}. Furthermore, recent
experimental advances in cold-atom systems enable realizations of the XXZ
chain and preparation of certain initial states \cite%
{Fukuhara2013,Jepsen2020}, providing an ideal platform for studying
nonequilibrium quantum dynamics.\ Specifically, the discovery of highly
excited many-body eigenstates of the XXZ Heisenberg model, referred to as
Bethe phantom states, has received much attention from both theoretical \cite%
{popkov2016obtaining,popkov2017solution,popkov2020exact,popkov2021phantom}
and experimental approaches \cite%
{Jepsen2020,jepsen2021transverse,hild2014far,jepsen2022long}. Previous work has examined the issue of many-body localization (MBL) arising
from strong disorder in a generalized XXZ model \cite{Stagraczynski2017}.
Here, our primary focus is on investigating whether the helix state can
become a quantum scar in the XXZ model in the absence of MBL. Although the
XXZ chain is an integrable system, the\ helix states are not supported by
the symmetry of the model \cite{popkov2021phantom}. Therefore, they are
regarded as a candidate of quantum scars in the presence of perturbation.
Importantly, a helix state essentially supports macroscopic long-range order
in an ultracold atom quantum simulator \cite{jepsen2022long}. However, it is
still an open question what happens if the XXZ chain\ becomes nonintegrable
in the presence of random field.

In this work, we investigate the influence of the external fields on the
stability of spin helix states in an XXZ Heisenberg model. We consider two
types of fields, (i) uniform field in $z$\ direction; (ii) random transverse
fields in $x$\ and $y$\ directions. These two fields are shown to play
different roles on the dynamics of a static helix state, which is the
eigenstate of an unperturbed XXZ Heisenberg model. Exact diagonalization on
a finite system shows that random transverse fields in $x$ and $y$
directions drive the transition from integrability to nonintegrability. The
energy level statistics is a Poisson distribution for weak random field, but
becomes Wigner--Dyson from Gaussian Unitary Ensemble (WD-GUE) distribution
as the strength of the random field increases. In the absence of the uniform
$z$ field, the random field causes the deviation of the evolved state from
the original helix state. However, in the presence of uniform $z$ field, the
static helix state becomes a dynamic helix state with a relatively long life
as a quantum scar state. We analyze the underlying mechanism in the
framework of perturbation method.\ It manifests that the uniform field can
suppress the effect of the random field. Numerical simulations for the
dynamics of a helix state show that (i) in the presence of uniform field, a
helix state exhibits a near perfect revival as quantum scars; (ii) in the
absence of uniform field, although the helix state deviates from being an
eigenstate, it still maintains a profile of a helix state within a long time.

The rest of this paper is organized as follows: In Sec.~\ref{Model
Hamiltonian and scar towers}, we introduce the model Hamiltonian and the
helix states arising from towers. With these preparations, in Sec.~\ref%
{Static and dynamic helix states} we present two types of helix states
arising from towers. We demonstrate the phase diagram of the system in Sec.~%
\ref{Nonintegrability} by means of statistics of energy levels. Sec.~\ref%
{Quntum scars and stable helix states} contributes to the dynamics of the
helix states in the nonintegrable system.\ Sec.~\ref{Summary} concludes this
paper.

\section{Model Hamiltonian and precise towers}

\label{Model Hamiltonian and scar towers}

We begin this section by introducing a Hamiltonian
\begin{equation}
H=H_{0}+H_{\mathrm{ran}}  \label{H_spin}
\end{equation}%
which consists of two parts. The unperturbed system

\[
H_{0}=\sum_{j=1}^{N}[s_{j}^{x}s_{j+1}^{x}+s_{j}^{y}s_{j+1}^{y}+\cos
q(s_{j}^{z}s_{j+1}^{z}-\frac{1}{4})+hs_{j}^{z}]
\]%
describes quantum spin XXZ Heisenberg chain with resonant\ anisotropy via
wave vector $q=2\pi n/N$ ($n\in \left[ 0,N-1\right] $), in a uniform field
in the $z$ direction, and the perturbation term%
\begin{equation}
H_{\mathrm{ran}}=\sum_{j}\left( x_{j}s_{j}^{x}+y_{j}s_{j}^{y}\right) ,
\end{equation}%
is the subjected random fields along the $x$\ and $y$-direction. Here the
field distribution is $x_{j}=$ran($-x,x$) and $y_{j}=$ran($-y,y$), where ran(%
$-b,b$) denotes a uniform random number within ($-b,b$). Here $%
s_{j}^{\lambda }$ ($\lambda =x,y,z$) are canonical spin-$1/2$ variables, and
obey the periodic boundary condition $s_{N+1}^{\lambda }\equiv
s_{1}^{\lambda }$.

We start with the case where $x=y=0$, which serves as the basis for the rest
of the study. We introduce a set of $q$-dependent spin operators
\[
s_{q}^{+}=\left( s_{q}^{-}\right) ^{\dag
}=\sum_{j=1}^{N}e^{iqj}s_{j}^{+},s^{z}=\sum_{j=1}^{N}s_{j}^{z},
\]%
which not surprisingly satisfy the Lie algebra commutation relations

\begin{equation}
\left[ s_{q}^{+},s_{q}^{-}\right] =2s^{z},\left[ s^{z},s_{q}^{\pm }\right]
=\pm s_{q}^{\pm }.
\end{equation}%
Unlike the isotropic case with $q=0$, $H_{0}$\ does not have SU(2) symmetry.
However, when we consider a subspace $W$ spanned by a set of states $\left\{
\left\vert \psi _{n}\right\rangle \right\} $ ($n\in \left[ 0,N\right] $),
which is defined as%
\begin{equation}
\left\vert \psi _{n}\right\rangle =\frac{1}{\Omega _{n}}\left(
s_{q}^{+}\right) ^{n}\left\vert \Downarrow \right\rangle .
\end{equation}%
Here the normalization factor $\Omega _{n}=\left( n!\right) \sqrt{C_{N}^{n}}$%
,\ and then we have $\left\vert \psi _{0}\right\rangle =$ $\left\vert
\Downarrow \right\rangle $ $=\prod_{j=1}^{N}\left\vert \downarrow
\right\rangle _{j}$ and $\left\vert \psi _{N}\right\rangle =$ $e^{iq\left(
1+N\right) N/2}\left\vert \Uparrow \right\rangle $, with $\left\vert
\Uparrow \right\rangle =\prod\nolimits_{j=1}^{N}\left\vert \uparrow
\right\rangle _{j}$,\ with $s_{j}^{z}\left\vert \downarrow \right\rangle
_{j}=-1/2\left\vert \downarrow \right\rangle _{j}$ ($s_{j}^{z}\left\vert
\uparrow \right\rangle _{j}=1/2\left\vert \uparrow \right\rangle _{j}$).
Importantly, straightforward derivation shows that $\left\{ \left\vert \psi
_{n}\right\rangle \right\} $\ is a set of eigenstates of $H_{0}$, i.e.,%
\begin{equation}
H_{0}\left\vert \psi _{n}\right\rangle =\left( n-N/2\right) h\left\vert \psi
_{n}\right\rangle .
\end{equation}%
Remarkably, it ensures that%
\begin{equation}
\{[H_{0},s_{q}^{\pm }]\mp hs_{q}^{\pm }\}W=0,
\end{equation}%
which indicates that $\left\{ \left\vert \psi _{n}\right\rangle \right\} $\
is a precise tower of energy according to the theorem in Ref. \cite%
{Daniel2020}. In parallel, state%
\begin{equation}
\left\vert \overline{\psi }_{n}\right\rangle =\frac{1}{\Omega _{n}}\left(
s_{-q}^{+}\right) ^{n}\left\vert \Downarrow \right\rangle ,
\end{equation}%
is also eigenstates of $H_{0}$, which share the same energy with $\left\vert
\psi _{n}\right\rangle $. Studies have shown that quantum scars can
be generated through towers \cite{Iversen2023,Kuno2020,Moudgalya2020}.
Additionally, research indicates that introducing perturbations into the
Heisenberg model can transition the system from integrable to chaotic, while
the towers persist \cite{Zhang2023}. Inspired by this, if after introducing
perturbations, this model can also become chaotic and the towers survive,
the towers will be able to generate quantum scars. We will discuss this in
detail in Secs.~\ref{Nonintegrability} and~\ref{Quntum scars and stable
helix states}.

\begin{figure*}[t]
\begin{center}
\includegraphics[bb=0 0 410 300,width=0.3\textwidth,clip]{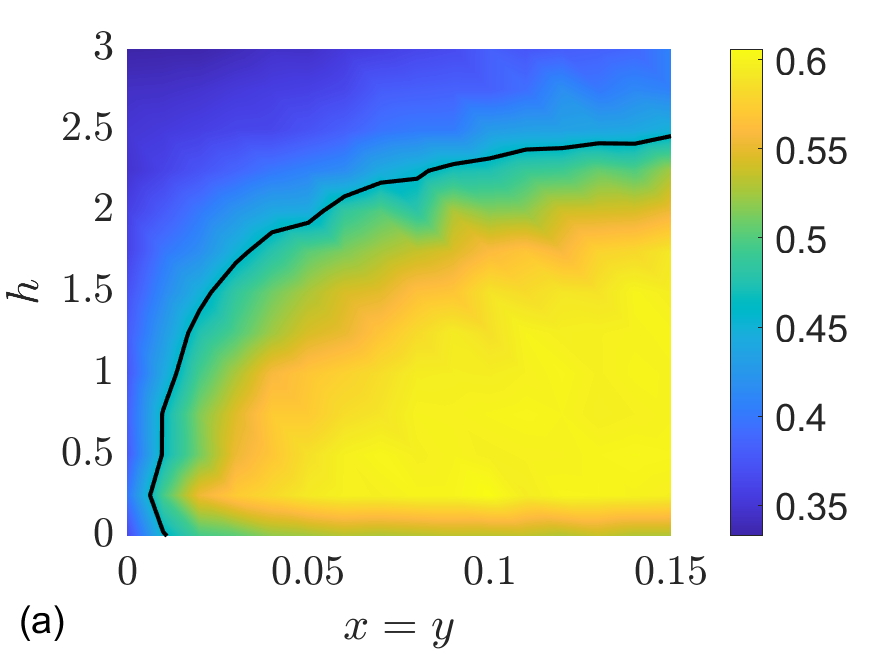} %
\includegraphics[bb=0 0 400 300,width=0.3\textwidth,clip]{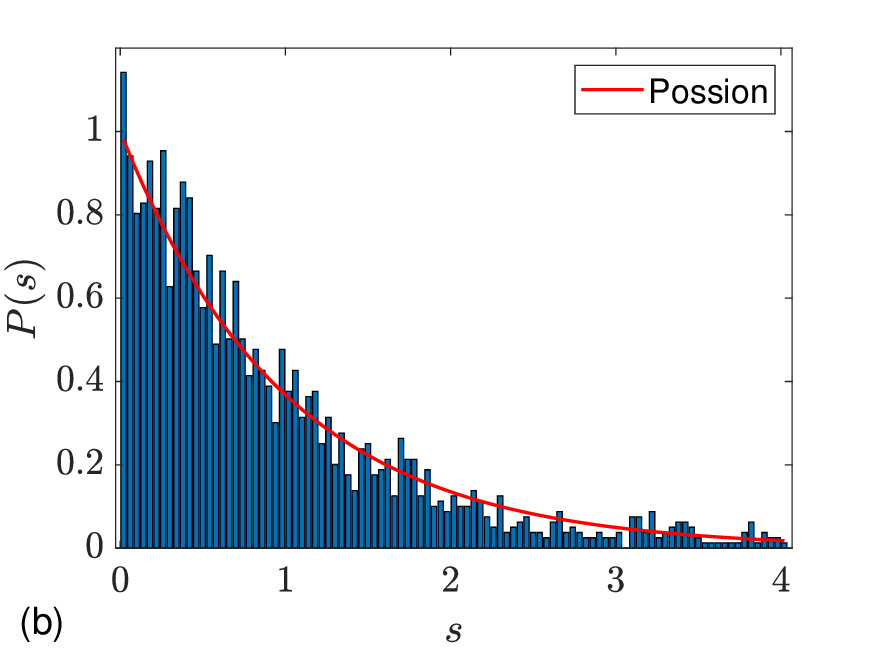} %
\includegraphics[bb=0 0 400 300,width=0.3\textwidth,clip]{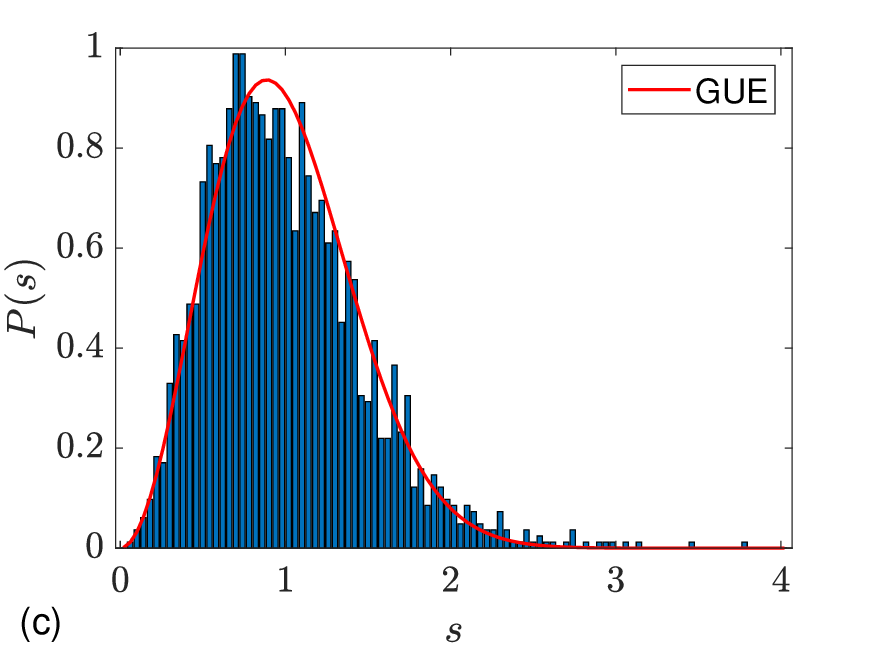} %
\end{center}
\caption{Exact diagonalization results on $r$-values and\ statistics of
energy level spacings for the model on an $N=12$ chain in Eq. (\protect\ref%
{H_spin}) with representative parameters. The results are obtained from the
average over $10$ sets of random number. (a) Color contour plot of the
average level spacing ratio $r$ as a function of $h$ and $x$. There are two
regions separated by the black contour line. The distribution of $r(h,x)$
indicates the phase diagram: The blue region represents the integrable phase
and the yellow region represents the nonintegrable phase. (b, c) Plots of $%
P(s)$ for two typical cases: $x=y=0.0$ and $x=y=0.05$, respectively. In both
cases, the field\ $h=0.5$. In order to avoid the unexpected degeneracy, the
coupling strength for the term $s_{1}^{x}s_{N}^{x}$ is taken as $0.85$. The
red lines indicate Poisson and WD-GUE distributions for comparison, the
characteristic of integrable and chaotic systems described by random matrix
theory. As expected, the distributions agree with the phase diagram in (a).}
\label{figure1}
\end{figure*}

\section{Static and dynamic helix states}

\label{Static and dynamic helix states}

In this section, we will introduce two types of helix states based on the
eigenstates of $H_{0}$. A generalized helix state is defined by a state in
the form%
\begin{equation}
\left\vert \Phi \right\rangle =\prod_{j=1}^{N}\left[ \cos \left( \theta
/2\right) \left\vert \uparrow \right\rangle _{j}+e^{i\Lambda (j,t)}\sin
\left( \theta /2\right) \left\vert \downarrow \right\rangle _{j}\right] ,
\end{equation}%
which is obviously an unentangled state. It represents the tensor
product of all spin precession states, forming a standard helix state. This
is a physically interesting state that has recently been studied both
theoretically and experimentally \cite%
{popkov2016obtaining,popkov2017solution,popkov2020exact,popkov2021phantom,jepsen2022long}.
So far, the study of this state mainly focuses on a simple case with $%
\Lambda (j,t)=qj+\omega t$. State $\left\vert \Phi \right\rangle $\ with
zero $\omega $\ is the eigenstate of a fine-tuned XXZ Heisenberg ring with
zero extrnal field. Here $\theta $\ is an arbitrary parameter, resulting in
a set of degenerate states of the system. It allows a multiple occupation of
a single magnon mode with nonzero wave vector $q$, as an analog of a
Goldstone mode in the anisotropic Heisenberg model \cite{popkov2021phantom}.
In this work, we focus on the case with nonzero $\omega $, arising from
nonzero external field. We introduce a local vector $\mathbf{h}_{l}=\left(
h_{l}^{x},h_{l}^{y},h_{l}^{z}\right) $ with $h_{l}^{\alpha }=\left\langle
\psi \right\vert s_{l}^{\alpha }\left\vert \psi \right\rangle $ ($\alpha
=x,y,z$) to characterize the helicity of a given state $\left\vert \psi
\right\rangle $.\ For eigenstates $\left\vert \psi _{n}\right\rangle $,
straightforward derivation of $h_{l}^{\alpha }(n)=\left\langle \psi
_{n}\right\vert s_{l}^{\alpha }\left\vert \psi _{n}\right\rangle $\ show
that
\[
h_{l}^{x}(n)=h_{l}^{y}(n)=0,h_{l}^{z}(n)=\frac{n}{N}-\frac{1}{2},
\]%
which is uniform, indicating that $\left\vert \psi _{n}\right\rangle $\ is
not a helix state. Nevertheless, in the following we will show that their
superposition can be helix states. These states can be classified into two
types of helix states: static and dynamic helix states.

\subsection{Static helix states}

We consider a superposition eigenstates in the form
\begin{equation}
\left\vert \phi (\theta )\right\rangle =\sum_{n}d_{n}\left\vert \psi
_{n}\right\rangle ,  \label{state}
\end{equation}%
where%
\begin{equation}
d_{n}=\sqrt{C_{N}^{n}}\left( -i\right) ^{n}\sin ^{n}\left( \theta /2\right)
\cos ^{\left( N-n\right) }\left( \theta /2\right) .
\end{equation}%
The corresponding helix vector is
\begin{equation}
\mathbf{h}_{l}=\frac{1}{2}[\sin \theta \sin \left( ql\right) ,\sin \theta
\cos \left( ql\right) ,-\cos \theta ],
\end{equation}%
which indicates that $\left\vert \phi (\theta )\right\rangle $\ is a helix
state for nonzero $\sin \theta $. Here $\theta $ is an arbitrary angle and
determines the profile of the state. This can be obtained easy when we
express it in the form.

\begin{equation}
\left\vert \phi (\theta )\right\rangle =\prod_{j=1}^{N}[\cos \left( \theta
/2\right) \left\vert \downarrow \right\rangle _{j}-ie^{iqj}\sin \left(
\theta /2\right) \left\vert \uparrow \right\rangle _{j}].  \label{phi}
\end{equation}%
It represents a tensor product of the precession states of all spins, which
is an unentangled state. It accords with the result $\left\vert \mathbf{h}%
_{l}\right\vert ^{2}=1/4$. The state $\left\vert \phi (\theta )\right\rangle
$\ is referred to as a static helix state since it is time-independent. In
parallel state
\begin{equation}
\left\vert \overline{\phi }(\theta )\right\rangle =\prod_{j=1}^{N}[\cos
\left( \theta /2\right) \left\vert \downarrow \right\rangle
_{j}-ie^{-iqj}\sin \left( \theta /2\right) \left\vert \uparrow \right\rangle
_{j}],
\end{equation}%
is another one with opposite helicity. We note that both $\left\vert \phi
(\theta )\right\rangle $\ and $\left\vert \overline{\phi }(\theta
)\right\rangle $\ are also eigenstates of $H_{0}$\ with zero $h$. Some
studies have already explored the dynamic generation of helix states driven
by local non-Hermitian fields in the XXZ Heisenberg model with the
Dzyaloshinskii-Moriya interaction \cite{Shi2023, Ma2022}, and now we are
investigating how these helix states evolve when an external field is
applied.

\begin{figure*}[t]
\begin{center}
\includegraphics[bb=0 0 400 320,width=0.3\textwidth,clip]{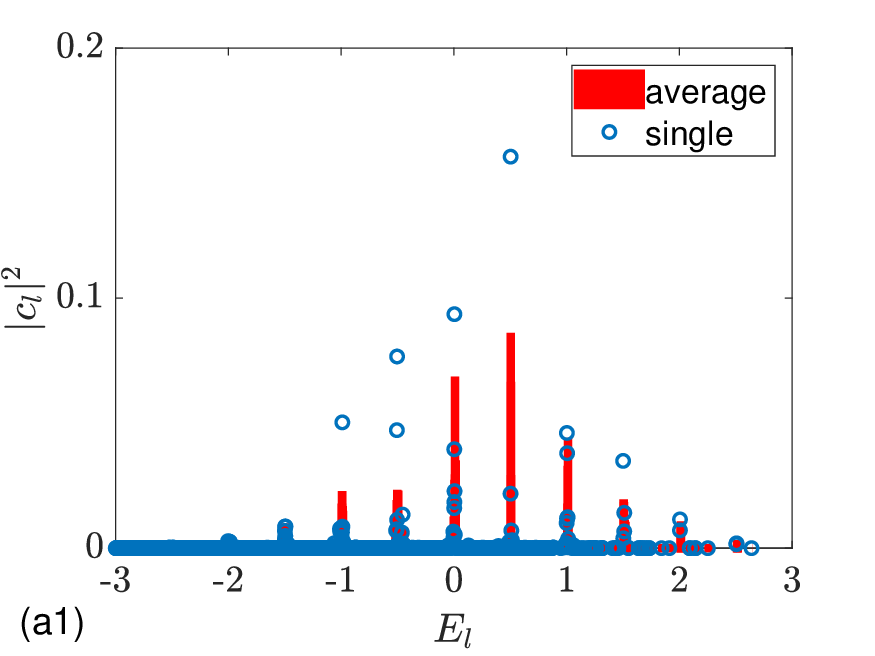} %
\includegraphics[bb=0 0 400 320,width=0.3\textwidth,clip]{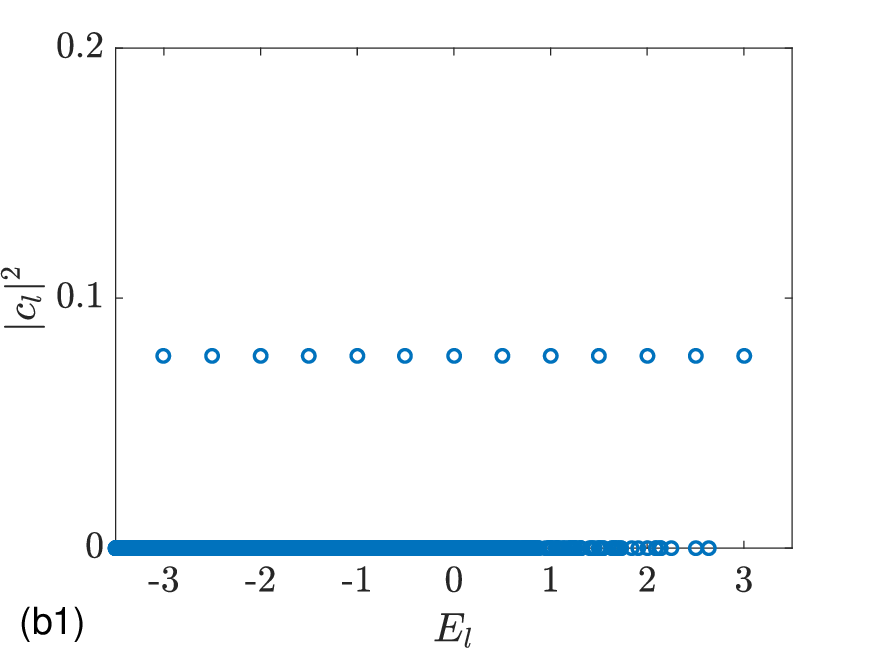} %
\includegraphics[bb=0 0 400 320,width=0.3\textwidth,clip]{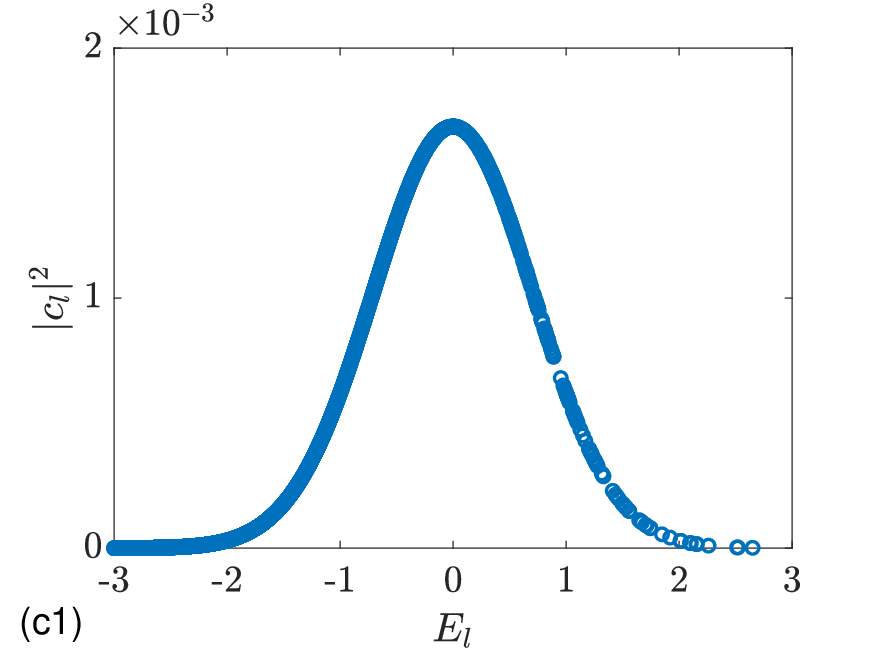} %
\includegraphics[bb=0 0 400 320,width=0.3\textwidth,clip]{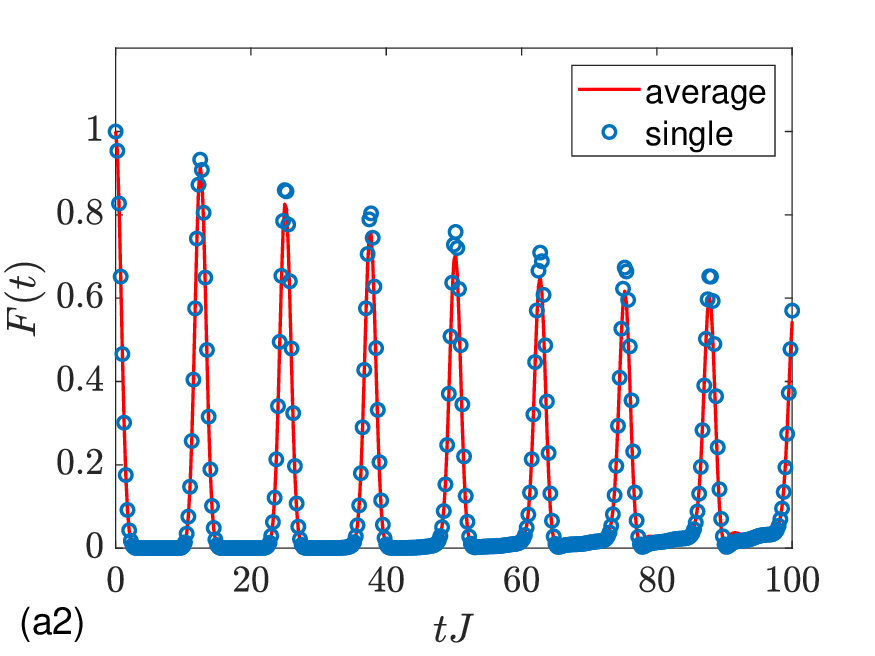} %
\includegraphics[bb=0 0 400 320,width=0.3\textwidth,clip]{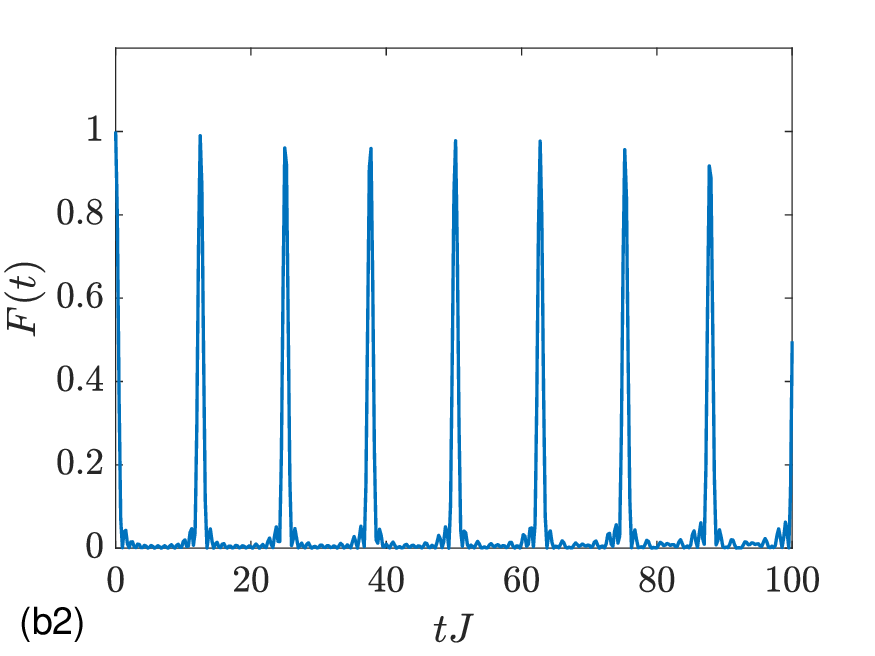} %
\includegraphics[bb=0 0 400 320,width=0.3\textwidth,clip]{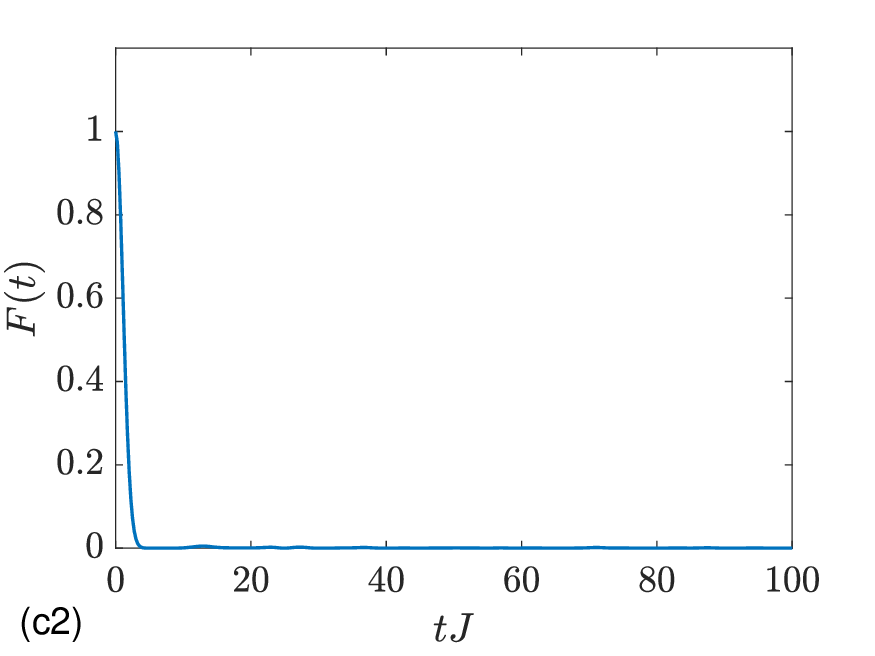}
\end{center}
\caption{Plots of the fidelity from Eq. (\protect\ref{fidelity}) for three
typical initial states. The parameters are $N=12$, $x=y=0.05$ and $h=0.5$.
The top panel displays the distribution of expansion coefficients $%
\left\vert c_{l}\right\vert ^{2}$ across energy level. The bottom panel
displays the fidelity. (a1), (a2) For the helix state. The red bar and red
line are obtained from the average over 20 sets of random numbers. (b1),
(b2) For a linear superposition of quantum states $\left\vert \protect%
\varphi _{l}\right\rangle $, which has the maximum overlap with quantum
state $\left\vert \protect\psi _{n}\right\rangle $. (c1), (c2) For an
initial state with a Gaussian distribution. We find that the helix state
exhibits periodic revival phenomena, whereas the Gaussian state does not.}
\label{figure2}
\end{figure*}

\subsection{Dynamic helix states}

When the external filed $h$\ switches on, $\left\vert \phi (\theta
)\right\rangle $\ and $\left\vert \overline{\phi }(\theta )\right\rangle $\
are no longer the eigenstates of $H_{0}$. Considering $\left\vert \phi
(\theta )\right\rangle $ as initial state at $t=0$, the evolved state is%
\begin{eqnarray}
\left\vert \phi _{0}(\theta ,t)\right\rangle
&=e^{-iE_{0}t}\sum_{n}e^{-inht}d_{n}\left\vert \psi _{n}\right\rangle
\nonumber \\
&=e^{-iE_{0}t}\times \prod_{j=1}^{N}[\cos \left( \theta /2\right) \left\vert
\downarrow \right\rangle _{j}-ie^{i\left( qj-ht\right) }\sin \left( \theta
/2\right) \left\vert \uparrow \right\rangle _{j}],
\end{eqnarray}%
where $E_{0}=-Nh/2$. The corresponding helix vector is
\begin{equation}
\mathbf{h}_{l}=\frac{1}{2}[\sin \theta \sin \left( ql-ht\right) ,\sin \theta
\cos \left( ql-ht\right) ,-\cos \theta ],
\end{equation}%
which satisfies the travelling-wave relation%
\begin{equation}
\mathbf{h}_{l}(t)=\mathbf{h}_{l-ht/q}(0).
\end{equation}%
This indicates that $\left\vert \psi \left( t\right) \right\rangle $\ acts
as a travelling wave with phase velocity $h/q$, then is referred to as
dynamic helix state. Obviously, it\ is also a perfect periodic function of
time, and then the system exhibits a perfect revival. To demonstrate this
phenomenon, plots of $h_{l}$ for several typical cases are presented in Fig. %
\ref{figure4}.

\begin{figure*}[t]
\begin{center}
\includegraphics[bb=0 0 400 320,width=0.3\textwidth,clip]{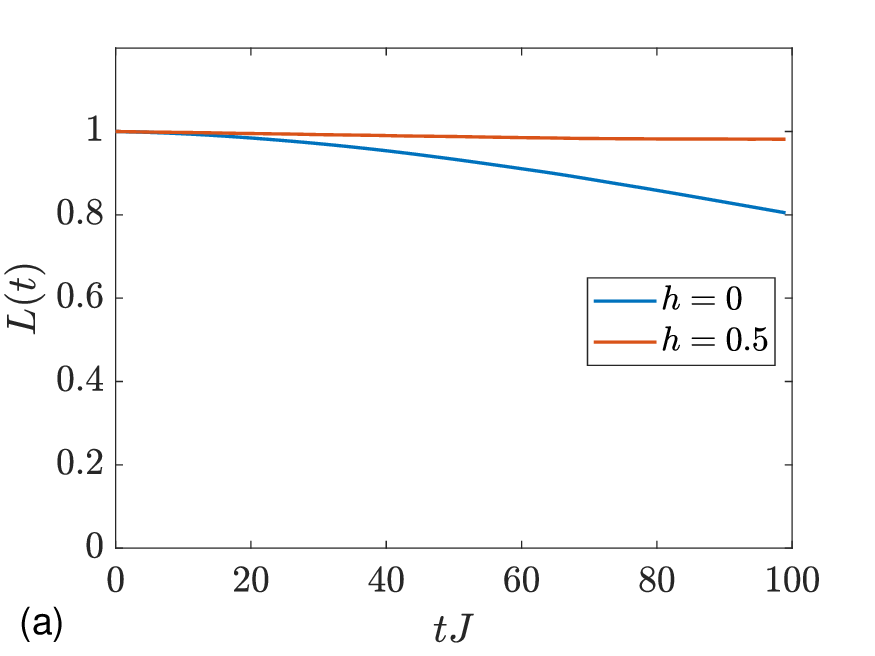} %
\includegraphics[bb=0 0 400 320,width=0.3\textwidth,clip]{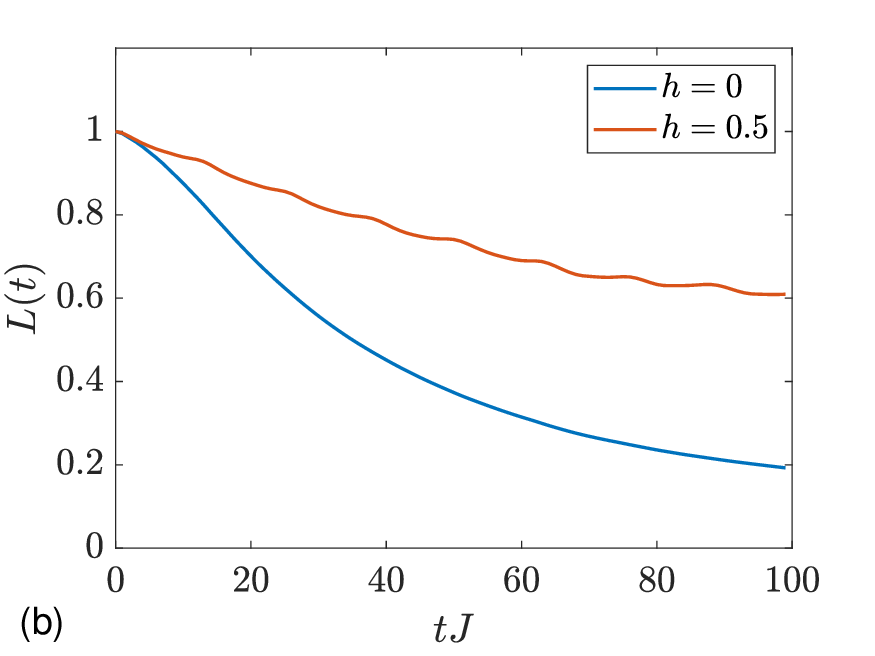} %
\includegraphics[bb=0 0 400 320,width=0.3\textwidth,clip]{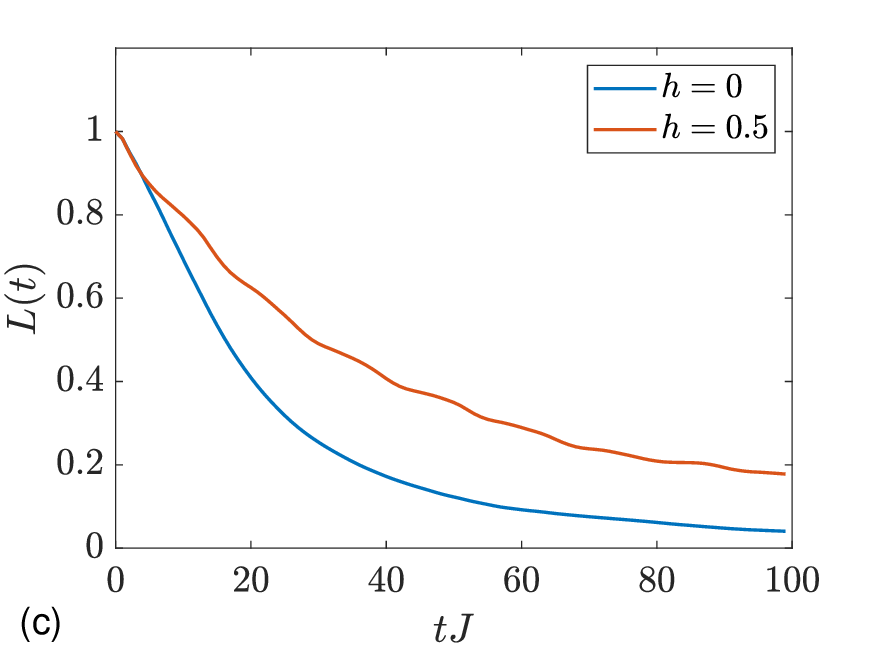}
\end{center}
\caption{Plots of the Loschmidt echo from (\protect\ref{Loschmidt_echo}) at
different disorder strength, where $N=12$, $h=0,0.5.$ (a) For $x=y=0.01.$
(b) For $x=y=0.05.$ (c) For $x=y=0.1.$ We find that the increase in disorder
strength will intensify the decay of the Loschmidt echo, while increasing
the magnetic field strength will suppress the decay of the Loschmidt echo.}
\label{figure3}
\end{figure*}

\section{Level statistics}

\label{Nonintegrability}

A popular approach to distinguishing between integrable and nonintegrable
model involves studying spectral statistics using tools derived from random
matrix theory. For a nonintegrable model, the statistical distribution of
level spacings follows the Wigner-Dyson distribution, while integrable
models follow the Poisson distribution. This distribution can also be
directly detected using the average level spacing ratio $r$-value \cite%
{Oganesyan2007, Atas2013}. which is the average of
\begin{equation}
r_{l}=\frac{\mathrm{min}\{s_{l},s_{l+1}\}}{\mathrm{max}\{s_{l},s_{l+1}\}},
\end{equation}%
where $s_{l}$ is level spacings $E_{l}-E_{l-1}$ for a selected set of energy
levels $\left\{ E_{l}\right\} $. $r\approx 0.53$ and $r\approx 0.60$ for
Wigner--Dyson from Gaussian Orthogonal Ensemble (WD-GOE) and WD-GUE
distributions, respectively, and $r\approx 0.39$ for the Poisson
distribution. This has been used in several works \cite%
{Stagraczynski2017,Zhang2023,Laumann2014,Schecter2019,Dong2023}. The XXZ
Heisenberg model is known to be integrable by Bethe ansatz methods \cite%
{Mikeska2004}. However, in the presence of $H_{\mathrm{ran}}$, we will show
the integrability of the system is destroyed based on numerical results for
a finite-size system.{\ In Fig. \ref{figure1}, we depict a phase diagram
illustrating the average level spacing ratio $r$ plotted against $h$ vs
disorder strength in the $x$ plane. }As evident from the figure, a
transition from integrable to nonintegrable phases is observed.

On the other hand, we also computed the statistical distribution of the
spacings $s$ between energy levels, denoted as $P(s)$. We start with the
case $x=y=0$, but arbitrary $h$. The distribution $P(s)$\ in each sector
indexed by the spin component in $z$\ direction is anticipated to follow a
Poisson distribution. Nevertheless, results obtained from exact
diagonalization of a small-sized system reveal that a near-perfect Poisson
distribution can be achieved when the translational symmetry is broken by a
slight deviation in the coupling strength of a single dimer. In the scenario
where $x$ and $y$ are nonzero, states in different sectors become
hybridized, necessitating the calculation of $P(s)$ across all levels. In
Fig. \ref{figure1}, we illustrate the level spacing statistics of the model
with a finite $N$. We take $x=y$\ for simplicity, but with different random
number series $\left\{ x_{j}\right\} $ and $\left\{ y_{j}\right\} $. We\
find that the level spacing distribution depends on the parameters $\left(
x,h\right) $: (i) $\left( x,h\right) $ $=\left( 0,0.5\right) $, $P(s)$\
follows the Poisson distribution as expected; (ii) $\left( x,h\right) $ $%
=\left( 0.05,0.5\right) $,\ $P(s)$ follows the WD-GUE distribution.

\section{Quantum scars and stable helix states}

\label{Quntum scars and stable helix states}

\begin{figure*}[t]
\begin{center}
\includegraphics[bb=0 0 150 105,width=0.3\textwidth,clip]{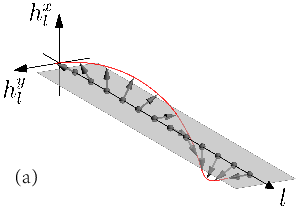} %
\includegraphics[bb=0 0 150 105,width=0.3\textwidth,clip]{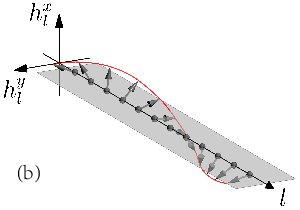} %
\includegraphics[bb=0 0 150 105,width=0.3\textwidth,clip]{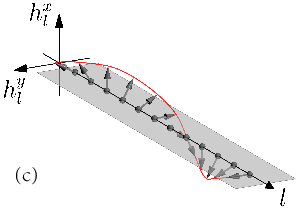}
\end{center}
\caption{Plots of the profile of the time evolution for the helix state in
Eq. (\protect\ref{phi}) with $\protect\theta =\protect\pi /2$\ as the
initial state under the Hamiltonian $H$\ with $x=y=0.05$ and $h=0$. (a) For $%
tJ=0.$ (b) For $tJ=20.$ (c) For $tJ=100.$ The corresponding Loschmidt echo $%
L(t)$\ is plotted Fig. \protect\ref{figure3}(b). It indicates that evolved
states are apparently still helix states with a slight deformation although
the fidelity becomes lower. The explanation is given in the text.}
\label{figure4}
\end{figure*}

As excited states, the energy levels for the subspace $\left\{ \left\vert
\psi _{n}\right\rangle \right\} $ and the helix state merge into the dense
spectrum.\ It has been proposed that the subspace $\left\{ \left\vert \psi
_{n}\right\rangle \right\} $\ is not supported by a symmetry of the
Hamiltonian $H_{0}$\ \cite{popkov2021phantom}. Therefore, as long as the
tower survive in the presence of $H_{\mathrm{ran}} $, even if it becomes a
quasi-tower, it is a candidate for quantum scar of the perturbed Hamiltonian
$H$. In this section, we focus on two questions by analytical analysis
approximately and numerical simulations: (i) Is a helix state immune to the
thermalization in the presence of $H_{\mathrm{ran}}$? (ii) Is the $\left\{
\left\vert \psi _{n}\right\rangle \right\} $ really a quantum scar? In the
following, we will demonstrate that the external field $h$\ can enhance the
stability of helix states, while $\left\{ \left\vert \psi _{n}\right\rangle
\right\} $\ can be a quantum scar even with the zero $h$ field.

We start with the estimation of the effect of the random field in $x$
direction on the dynamics of a helix state by considering a simplified
perturbation field%
\begin{equation}
H_{\mathrm{ran}}=x_{0}s_{l}^{x},
\end{equation}%
which is a local field in $x$ direction at site $l$. The matrix
representation of Hamiltonian $H$ in the subspace $\left\{ \left\vert \psi
_{n}\right\rangle \right\} $\ is an $\left( N+1\right) \times \left(
N+1\right) $ matrix $M$ with nonzero matrix elements

\begin{equation}
\left( M\right) _{n,n+1}=\left( M\right) _{n+1,n}^{\ast }=e^{iql}\frac{x_{0}%
}{2N}\sqrt{n\left( N+1-n\right) },
\end{equation}%
with $n=\left[ 1,N\right] $, and%
\begin{equation}
\left( M\right) _{n,n}=(n-1-N/2)h,
\end{equation}%
with $n=\left[ 1,N+1\right] $. It is obviously solvable matrix with equal
spacing energy levels, i.e.%
\begin{equation}
E_{n}=(n-1-N/2)\sqrt{h^{2}+\left( \frac{x_{0}}{N}\right) ^{2}},
\end{equation}%
with $n=\left[ 1,N+1\right] $. In the case with zero $h$, we have $%
E_{n}\approx \frac{x_{0}}{N}(n-1-N/2)$, while $E_{n}\approx $ $(n-1-N/2)h$\
in the case with $h^{2}\gg \left( x_{0}/N\right) ^{2}$. From the perspective
of energy correction, the effect of $H_{\mathrm{ran}}$\ becomes weak in the
presence of field $h$. This implies that $H_{\mathrm{ran}}$\ may damage the
helix state even worse in the zero $h$ field. It suggest that one can
suppress the thermalization on the subspace $\left\{ \left\vert \psi
_{n}\right\rangle \right\} $\ by $h$ field.

Now we demonstrate and verify our prediction by numerical simulation on the
dynamic processes. Our major aspects of concerns are (i) the revival of the
helix state in comparison with a non-scarring state; (ii) the deviation of
the evolved state in the presence of random field. Firstly, we consider the
time evolutions for three types of initial states, which are expressed in
the form
\begin{equation}
\left\vert \psi \left( 0\right) \right\rangle =\sum_{l}c_{l}\left\vert
\varphi _{l}\right\rangle
\end{equation}%
where $\left\vert \varphi _{l}\right\rangle $\ is the eigenstate with energy
$E_{l}$\ of the perturbed Hamiltonian $H$, i.e., $H\left\vert \varphi
_{l}\right\rangle =E_{l}\left\vert \varphi _{l}\right\rangle $. Here $%
\left\{ E_{l},\left\vert \varphi _{l}\right\rangle \right\} $\ depends on
the generated random number series. Three types of $\left\{ c_{l}\right\} $\
are obtained from 1) $\left\vert \psi \left( 0\right) \right\rangle
=\left\vert \phi \left( \pi /2\right) \right\rangle $; 2) Let the quantum
state $\left\vert \psi \left( 0\right) \right\rangle $\ be a linear
superposition of quantum states $\left\vert \varphi _{l}\right\rangle $,
which has the maximum overlap with the quantum state $\left\vert \psi
_{n}\right\rangle $; 3) Gaussian distribution of $\left\{ c_{l}\right\} $.
We introduce the quantity%
\begin{equation}
F\left( t\right) =\left\vert \langle \psi \left( t\right) \left\vert \psi
\left( 0\right) \right\rangle \right\vert ^{2}  \label{fidelity}
\end{equation}%
to characterize the quality of the\ revival. The plots of $F\left( t\right) $%
\ in Fig. \ref{figure2} for several typical cases show that the revival
strongly depends on the initial state. The time evolution in \ref{figure2}%
(b1, b2) shows that the near perfect revival can be achieved even in the
nonintegrable system. It indicates that the tower remains stable
after the introduction of perturbation, and this long-lived oscillatory
dynamics of the helix state can serve as evidence for helix state being a
quantum scar. Secondly, we measure the influence of random and uniform field
on the helix state and explore methods to enhance the stability of the helix
state under perturbations. We consider the time evolutions of a same state
under two driven Hamiltonian. In this case, the Loschmidt echo defined by%
\begin{equation}
L\left( t\right) =\left\vert \langle \phi (\theta ,t)\left\vert \phi
_{0}(\theta ,t)\right\rangle \right\vert ^{2}  \label{Loschmidt_echo}
\end{equation}%
is usually employed, with $\left\vert \phi (\theta ,t)\right\rangle
=e^{-iHt}\left\vert \phi (\theta )\right\rangle $ and $\left\vert \phi
_{0}(\theta ,t)\right\rangle =e^{-iH_{0}t}\left\vert \phi (\theta
)\right\rangle $. In Fig. \ref{figure3}, we plot the Loschmidt echo obtained
from systems with different disorder strength and uniform field $h$.
Because $\left\vert \phi _{0}(\theta ,t)\right\rangle $\ can
maintain a perfect helix state, the Loschmidt echo can reflect the extent to
which the dynamic behavior deviates from the helix state under
perturbations. It is evident that the decay of Loschmidt echo accelerates
with increasing disorder strength. This is easily understood, but it is
worth noting that regardless of the level of disorder strength, introducing
a uniform magnetic field consistently suppresses the decay of Loschmidt
echo. This indicates that the uniform magnetic field can enhance the
stability of the helix state.

\begin{figure*}[t]
\begin{center}
\includegraphics[bb=0 0 400 320,width=0.4\textwidth,clip]{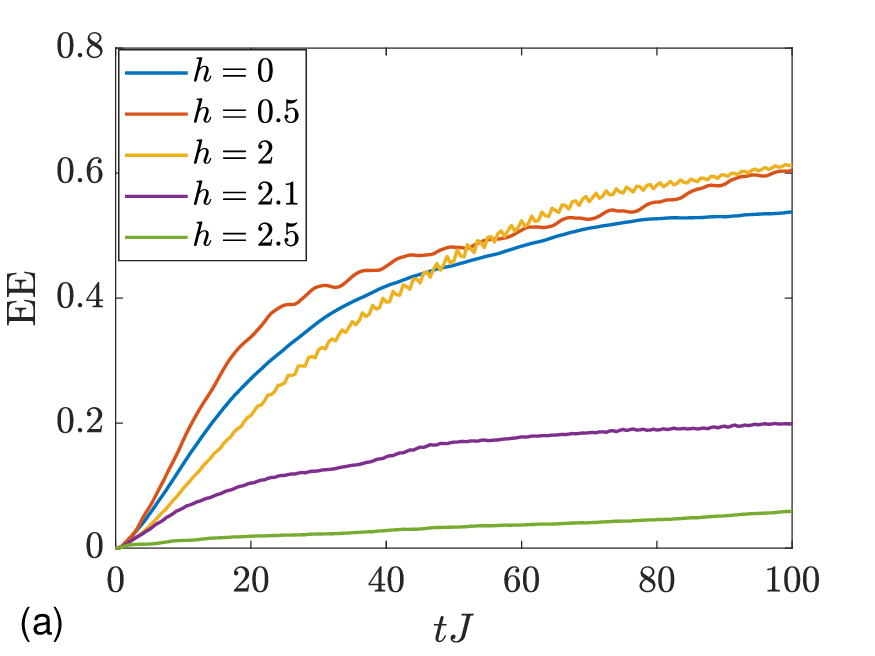} %
\includegraphics[bb=0 0 400 320,width=0.4\textwidth,clip]{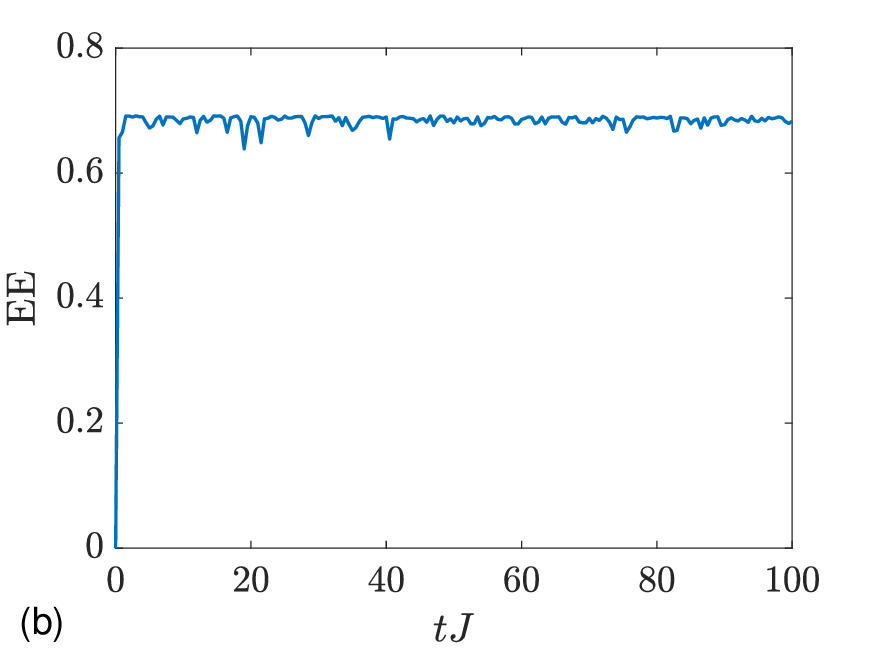}
\end{center}
\caption{Plots of the entanglement entropy for helix state and Neel state,
where $N=12$, $x=y=0.05$. (a) For the helix state with $h=0,0.5,2,2.1,2.5$.
(b) For the Neel state with $h=0.5$. All of them are obtained from the
average over 20 sets of random number. We find that when $h $ is less than
2, the influence on the entanglement entropy of the helix state is minimal.
However, when $h$ exceeds 2, the entanglement entropy is rapidly suppressed.
In either scenario, the growth of entanglement entropy of the helix state is
significantly slower than that of the Neel state.}
\label{figure5}
\end{figure*}

However, we would like to point out that the fast decay of Loschmidt echo\
does not indicate the collapse of the helix state, but may be only different
from the state $\left\vert \phi _{0}(\theta ,t)\right\rangle $. We
demonstrate this point by the plots of $\mathbf{h}_{l}$ for $\left\vert \phi
(\theta ,t)\right\rangle $ in Fig. \ref{figure4}. We compute the time
evolution for extreme case, in which the Loschmidt echo decays must faster.
Numerical results show that the profile of the helix state changes slightly.
This can be explained by the estimation on the Loschmidt echo of individual
spin when $L\left( t\right) =0.2$ and $0.6$, for example,$\ \sqrt[12]{0.2}%
=0.87$, $\sqrt[12]{0.6}=0.96$, which is similar to the orthogonality
catastrophe by Anderson \cite{Anderson1967,Anderson19672}.

Based on this, a favorable result of Loschmidt echo can indicate that the
scar behavior is indeed favorable and help people know how the scar evolves
in experiment, but a decay of Loschmidt echo does not necessarily imply poor
scar behavior. Therefore, we also investigated the entanglement entropy (EE)
of the helix state to measure its scar extent. The entanglement entropy is
defined as $-\mathrm{Tr}\left( \rho _{A}\log \rho _{A}\right) $, where A is
the reduced density matrix of the subsystem A. Due to symmetry
considerations, we select the first spin as the subsystem in this context.
The results of the entanglement entropy under various uniform fields $h$\
are depicted in Fig. \ref{figure5}(a). It can be observed that when the
uniform field $h$\ is less than 2, it has minimal impact on the entanglement
entropy. However, when $h$\ exceeds 2, the entanglement entropy rapidly
suppressed. This phenomenon is attributed to the system transitioning from
chaotic to integrable, as illustrated in Fig. \ref{figure1}(a). Since the
entanglement entropy of the Gaussian state is inherently high, which is
unfavorable for observing the evolution results, we plotted the entanglement
entropy of the Neel state, a commonly used quantum state in quantum
information, in Fig. \ref{figure5}(b). By comparison, it is evident that as
long as the system remains chaotic, regardless of the value of $h$, the
helix state can be considered a robust quantum scar.

\section{Conclusion}

\label{Summary}

In summary, we have investigated the influence of the external fields on the
stability of spin helix states in a XXZ Heisenberg model. Exact
diagonalization on a finite system shows that random transverse fields in $x$
and $y$ directions drive the transition from integrability to
nonintegrability.\ In such a system, the helix state can be regarded as
quantum scar. Simultaneously, the presence of uniform $z$\ field enables the
helix state to better maintain its dynamical nature, allowing for a clearer
understanding of its evolution behavior. However, the entanglement entropy
reveals that irrespective of the presence of uniform $z $\ field, as long as
the system remains chaotic, the scar extent of the helix state shows no
significant variation. It is expected to be insightful for quantum
engineering of a helix state in experiments.

\ack
We acknowledge the support of the National Natural Science Foundation of
China (Grants No. 11705127 and No. 11874225).

\end{document}